\documentclass[prl,aps,amsfonts,amssymb,amsmath,floats,floatfix,showpacs,twocolumn,superscriptaddress,unsortedaddress]{revtex4}
\usepackage{graphics}
\usepackage{epsfig}

\begin{document}
\title{Fermi Surface and Quasiparticle Excitations of overdoped Tl$_2$Ba$_2$CuO$_{6+\delta}$ by ARPES}
\author{M. Plat$\acute{{\rm e}}$}
\affiliation{Department of Physics {\rm {\&}} Astronomy, University of British Columbia, Vancouver, British Columbia, Canada V6T\,1Z1}
\author{J.D.F. Mottershead}
\affiliation{Department of Physics {\rm {\&}} Astronomy, University of British Columbia, Vancouver, British Columbia, Canada V6T\,1Z1}
\author{I.S. Elfimov}
\affiliation{Department of Physics {\rm {\&}} Astronomy, University of British Columbia, Vancouver, British Columbia, Canada V6T\,1Z1}
\author{D.C. Peets}
\affiliation{Department of Physics {\rm {\&}} Astronomy, University of British Columbia, Vancouver, British Columbia, Canada V6T\,1Z1}
\author{Ruixing Liang}
\affiliation{Department of Physics {\rm {\&}} Astronomy, University of British Columbia, Vancouver, British Columbia, Canada V6T\,1Z1}
\author{D.A. Bonn}
\affiliation{Department of Physics {\rm {\&}} Astronomy, University of British Columbia, Vancouver, British Columbia, Canada V6T\,1Z1}
\author{W.N. Hardy}
\affiliation{Department of Physics {\rm {\&}} Astronomy, University of British Columbia, Vancouver, British Columbia, Canada V6T\,1Z1}
\author{S. Chiuzbaian}
\affiliation{Swiss Light Source, Paul Scherrer Institut, CH-5234 Villigen, Switzerland}
\author{M. Falub}
\affiliation{Swiss Light Source, Paul Scherrer Institut, CH-5234 Villigen, Switzerland}
\author{M. Shi}
\affiliation{Swiss Light Source, Paul Scherrer Institut, CH-5234 Villigen, Switzerland}
\author{L. Patthey}
\affiliation{Swiss Light Source, Paul Scherrer Institut, CH-5234 Villigen, Switzerland}
\author{A. Damascelli}
\affiliation{Department of Physics {\rm {\&}} Astronomy, University of British Columbia, Vancouver, British Columbia, Canada V6T\,1Z1}

\begin{abstract}
The electronic structure of the high-$T_c$ superconductor Tl$_2$Ba$_2$CuO$_{6+\delta}$ is studied by ARPES. For a very overdoped $T_c\!=\!30$\,K
sample, the Fermi surface consists of a single large hole pocket centered at ($\pi$,$\pi$) and is approaching a topological transition. Although
a superconducting gap with $d_{x^2-y^2}$ symmetry is tentatively identified, the quasiparticle evolution with momentum and binding energy
exhibits a marked departure from the behavior observed in under and optimally doped cuprates. The relevance of these findings to scattering,
many-body, and quantum-critical phenomena is discussed.
\end{abstract}

\date{Received 5 November 2004}

\pacs{74.25.Jb, 74.72.Jt, 79.60.-i}

\maketitle

%
%
\vskip2pc

\narrowtext

Angle-resolved photoemission spectroscopy (ARPES) on the high-$T_c$ superconductors (HTSCs) has provided
crucial insights into the complex electronic structure of these materials. However, despite an intense
experimental and theoretical effort, no conclusive agreement has yet been reached on the interpretation of
some of the most fundamental results \cite{Andrea:RMP}. This is partly due to the fact that most of the
available ARPES data on cuprates have been obtained on a limited set of materials, such as
La$_{2-x}$Sr$_{x}$CuO$_4$ (LSCO) and the single and double CuO$_2$ layer Bi-cuprates (Bi2201 and Bi2212),
whose electronic structure is complicated by several materials issues. These include chemical and
particularly cation disorder, as well as family-specific problems: lattice distortions and spin/charge
instabilities in LSCO, superstructure modulations in the Bi-cuprates, and band splitting due to the presence
of CuO$_2$ bilayer blocks in Bi2212 \cite{Andrea:RMP}.

Important breakthroughs may come from the study by ARPES of the Tl-cuprates and in particular single layer
Tl$_2$Ba$_2$CuO$_{6+\delta}$ (Tl2201). Owing to a well-ordered crystal structure with very flat CuO$_2$
planes far apart from each other, its electronic structure should be free of many complications found in
other cuprates. At the same time, its $T_c^{max}\!\backsimeq\!93$\,K is one of the highest among single layer
materials ($T_c^{max}\!<\,40$\,K for Bi2201 and LSCO, possibly due structural effects or to larger and/or
more harmful cation disorder \cite{Eisaki}). Most importantly, Tl2201 can be synthesized over a wide doping
range extending from the optimal to the very overdoped regime. Although the latter is accessible in LSCO and
Pb-doped Bi-cuprates, these systems are affected by the cation disorder associated with Sr and Pb doping.
Thus Tl2201 offers a unique opportunity to reach the heavily overdoped side of the cuprate phase diagram,
where important hints of Fermi liquid behavior were obtained \cite{Proust}. As most of the research effort
has focused on the optimal/underdoped regime, in an attempt to understand the connection between Mott-Hubbard
insulating behavior and superconductivity, the study of heavily overdoped Tl2201 represents an important
alternative approach. At this stage, it has already been established that near optimal doping Tl2201 exhibits
a few key features common to most cuprates, such as a $d_{x^2-y^2}$ superconducting (SC) gap \cite{Tsuei} and
the ($\pi$,$\pi$) magnetic resonant mode \cite{He}. Additionally, in the very overdoped regime a coherent
three-dimensional Fermi surface (FS) was observed in angular magnetoresistance oscillations measurements
(AMRO) \cite{Hussey}.

Unfortunately, until now, ARPES experiments on Tl2201 have been severely hampered by the low-quality and/or
short lifetime of the cleaved surfaces, which resulted in poorly resolved spectroscopic features. In this
Letter we present the first extensive ARPES study of the low-energy electronic structure of overdoped Tl2201
crystals \cite{Peets}. These results provide us with detailed information on the FS and quasiparticle (QP)
dispersion. Although a SC gap consistent with the usual $d_{x^2-y^2}$ symmetry is observed, the ARPES
lineshapes exhibit an unexpected momentum dependence: contrary to the case of under and optimally-doped
cuprates \cite{Andrea:RMP,Zhou, Yoshida}, QPs are sharp near ($\pi$,0), i.e. the {\it antinodal} region where
the gap is maximum, and broad at ($\pi$/2,$\pi$/2), i.e. the {\it nodal} region where the gap vanishes. In
addition, while the QP linewidth at ($\pi$/2,$\pi$/2) increases as a function of binding energy, in the
($\pi$,0) region it is sharper at the bottom of the band than closer to the Fermi energy ($E_F$).

ARPES experiments were carried out at the Swiss Light Source on the SIS Beamline with a Scienta 2002
analyzer, circularly polarized 59\,eV photons, and energy/angular resolutions of
$\sim\!24$\,meV/$0.2^{\circ}$. Tl2201 single crystals were grown by a copper-rich self-flux method
\cite{Peets}, with stoichiometry Tl$_{1.88(1)}$Ba$_2$Cu$_{1.11(2)}$O$_{6+\delta}$ corresponding to Cu
substitution on the Tl site, thus away from the CuO$_2$ planes \cite{Shimakawa}. After careful annealing in
controlled oxygen partial pressure, overdoped samples with $T_c$ from 5 to 90\,K were obtained; their high
quality is evidenced by the narrow SC transitions, e.g. $\Delta T_c\!\simeq\!0.7$\,K for $T_c\!=\!67$\,K.
\begin{figure}[t!]
\centerline{\epsfig{figure=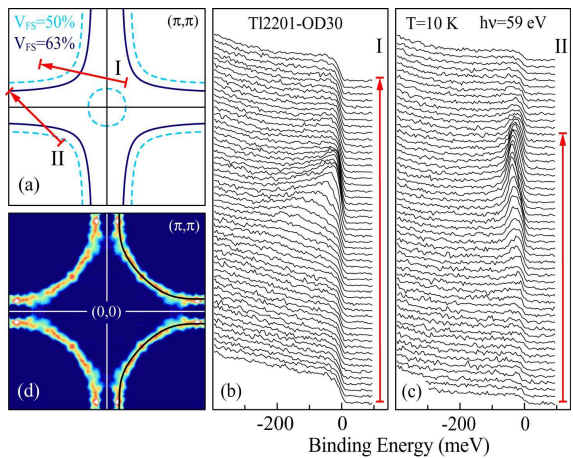,width=0.87\linewidth,clip=}} \vspace{-.1cm}
\caption{(color online). (a) LDA FS for two different doping levels corresponding to a volume, counting
holes, of 50\% (cyan, dashed) and 63\% (blue, solid) of the BZ. (b,c) ARPES spectra taken at $T\!=\!10$\,K on
Tl2201-OD30 along the directions marked by arrows in (a). (d) ARPES FS of Tl2201-OD30 along with a
tight-binding fit of the data (black lines).}  \label{fig1}
\end{figure}
The ARPES data were acquired in first and second Brillouin zones (BZ), returning analogous features, on two
overdoped samples with $T_c\!=\!63$ and 30\,K (Tl2201-OD63; Tl2201-OD30). The samples, cleaved at 10\,K and
$6\!\times\!10^{-11}$\,torr, were kept at 10\,K at all times.

Typical ARPES data from Tl2201-OD30 are presented in Fig.\,\ref{fig1} together with the FS obtained from our
band structure calculations within the local density approximation (LDA), which are in good agreement with
previous calculations \cite{Hamann,Singh}, and a tight-binding fit of the experimentally determined FS. The
spectra in Fig.\,\ref{fig1}b,c were measured along momentum space directions near the nodal and antinodal
regions of the BZ, as indicated by the arrows in Fig.\,\ref{fig1}a. Dispersive features are clearly
observable, with a behavior which is ubiquitous among the cuprates \cite{Andrea:RMP}. Close to the nodal
direction the QP peak exhibits a pronounced dispersion that can be followed over $\sim\!250$\,meV below
$E_F$; near ($\pi$,0), on the other hand, the band is much shallower with a van Hove singularity
$\sim\!39$\,meV below $E_F$. By integrating over a $\pm 5$\,meV window about $E_F$ the ARPES spectra
normalized at high binding energies, one obtains an estimate for the normal-state FS (Fig.\,\ref{fig1}d; the
$E_F$-intensity map across two BZ's was downfolded to the reduced zone scheme and symmetrized with respect to
the BZ diagonal, taking an average for equivalent $k$ points, and then fourfolded). As discussed later, at
$T\!=\!10$\,K a $d$-wave SC gap is open along the FS; thus this procedure returns the loci of minimum
excitation energy across the gap, which however correspond to the underlying normal-state FS crossings
\cite{Andrea:RMP}.

\begin{figure}[t!]
\centerline{\epsfig{figure=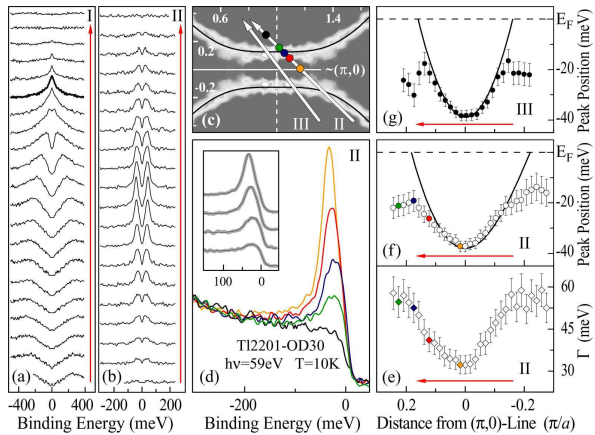,width=0.9\linewidth,clip=}} \vspace{-.1cm}
\caption{(color online). (a,b) Symmetrization of the ARPES spectra from along cut I and II in
Fig.\,\ref{fig1}. (c) Enlarged view of the FS of Tl2201-OD30 near ($\pi$,0). (d) Selected spectra from along
cut II in (c); their $k$-space positions are indicated by circles of corresponding color. (e,f) QP linewidth
$\Gamma$ and peak position from a Lorentzian fit of the energy distribution curves along cut II in (c). (g)
Similarly, QP peak position along cut III in (c). Black lines in (c,f,g) are the tight-binding results.}
\label{fig2}
\end{figure}

The FS of Tl2201-OD30 (Fig.\,\ref{fig1}d) consists of a large hole-pocket centered at ($\pi$,$\pi$) which, as
suggested by the low binding energy of the van Hove singularity (Fig.\,\ref{fig1}c), appears to be
approaching a topological transition from hole to electron-like. The FS volume, counting holes, is
63\,$\pm$\,2\% of the BZ corresponding to a carrier concentration of $1.26\pm0.04$\,hole/Cu atom, in very
good agreement with Hall-coefficient \cite{Mackenzie} and AMRO \cite{Hussey} experiments, which found 1.30
and 1.24 itinerant holes, respectively, in slightly more overdoped samples. These measurements all indicate
that the low-energy electronic structure of very overdoped Tl2201 is dominated by a single CuO band. In both
ARPES and AMRO data there is no evidence for the TlO band that in LDA calculations crosses $E_F$ and gives
rise to a small electron pocket centered at $k\!=\!(0,0)$ for non-oxygenated (i.e., $\delta\!=\!0$) Tl2201
(Fig.\,\ref{fig1}a, dashed FS). This however is no surprise even within the independent particle picture. In
fact, adjusting the chemical potential in the calculations in a rigid-band-like fashion to match the doping
level of our Tl2201-OD30 sample (as determined by the total FS volume), the TlO band is emptied of its
electrons and the LDA FS reduces to the single CuO pocket (Fig.\,\ref{fig1}a, solid FS). Since full depletion
of the TlO band takes place for $\Delta E_F\!\simeq\!-0.159$\,eV, corresponding to the removal of 0.024
electrons from the TlO band (as well as 0.109 from the CuO band), already the deviation of the Tl$^{3+}$ and
Cu$^{2+}$ content of our samples from the stoichiometric ratio 2:1, which contributes
$\sim\!0.14$\,hole/formula unit, would be sufficient to empty the TlO band even in the non-oxygenated
$\delta\!=\!0$ case. In this sense, the Tl-Cu non-stoichiometry and the presence of the TlO band cooperate in
pushing the $\delta\!=\!0$ system away from 1/2-filling, which may help explain why non-oxygenated Tl2201 is
not a charge transfer insulator like undoped (i.e., $x\!=\!0$) LSCO \cite{Singh}. As for the detailed shape
of the FS, which in LDA calculations is more square than in ARPES and AMRO experiments, better agreement
would require the inclusion in the calculations of correlation effects and/or O-doping beyond a rigid-band
picture. Alternatively the ARPES data can be modelled by the tight-binding dispersion $\epsilon_{\bf
k}\!=\!\mu\!+\!\frac{t_1}{2}(\cos k_x\!+\!\cos k_y)\!+\!t_2\cos k_x\cos k_y\!+\!\frac{t_3}{2}(\cos
2k_x\!+\!\cos 2k_y)\!+\!\frac{t_4}{2}(\cos 2k_x\cos k_y\!+\!\cos k_x\cos 2k_y)\!+\!t_5\cos 2k_x\cos 2k_y$, as
in Ref.\,\onlinecite{NormanPRB} (setting $a\!=\!1$ for the lattice constant). With parameters
$\mu\!=\!0.2438$, $t_1\!=\!-0.725$, $t_2\!=\!0.302$, $t_3\!=\!0.0159$, $t_4\!=\!-0.0805$, $t_5\!=\!0.0034$,
all expressed in eV, this dispersion reproduces both the FS shape (Fig.\,\ref{fig1}d) and the QP energy at
(0,0) and especially near ($\pi$,0) (Fig.\,\ref{fig2}f,g).

The analysis of the ARPES spectra in Fig.\,\ref{fig2} indicates a SC gap consistent with a $d_{x^2-y^2}$
form. Due to the lack of normal state data, the opening of the gap for this Tl2201-OD30 sample could not be
followed via the shift of the leading edge midpoint (LEM) across $T_c$, as is commonly done (this was however
possible in subsequent temperature dependent experiments on a less overdoped $T_c\!=\!74$\,K sample
\cite{Jeff}). In the present case the existence of a gap can be most easily visualized by the comparison of
nodal and antinodal symmetrized spectra \cite{NormanN}, in particular by the presence of a peak at $E_F$
along the nodal direction (signature of a FS crossing; bold line in Fig.\,\ref{fig2}a) and by the lack
thereof along the antinodal (Fig.\,\ref{fig2}b). For a more quantitative analysis, we performed a fit of the
spectra along different $k$-space cuts intersecting the underlying normal state FS (Fig.\,\ref{fig2}d; as
lineshape we used a Lorentzian QP peak plus a step-like background identified by the ARPES intensity at
$k\!\gg\!k_F$, all multiplied by a Fermi function and convoluted with the instrumental energy resolution
function \cite{Andrea:RMP}). As shown in Fig.\,\ref{fig2}f and \ref{fig2}g, where the fit results are
compared to our tight-binding dispersion for antinodal cuts II and III, the QP peak does not reach $E_F$ when
approaching $k_F$; instead it disperses back to higher binding energies losing intensity after having reached
a minimum value $\simeq\!17$\,meV. This behavior is a hallmark of Bogoliubov QPs, and reveals the opening of
a SC gap near ($\pi$,0). Due to the finite temperature and limited resolution of the experiment, the LEM of
the $k_F$ antinodal spectra is located at $\simeq\!2$\,meV above $E_F$. Our fitting procedure and detailed
simulations by Kordyuk {\it et al.} \cite{Kordyuk}, for comparable experimental parameters, suggest that the
observed location of the LEM is consistent with the presence of a SC gap $\Delta\!\simeq\!8$\,meV; this gap
value is a factor $\sim\!2$ smaller than the QP peak position, as empirically noted for most HTSCs
\cite{Andrea:RMP}.
Along cuts intersecting the FS at the nodes (not shown), the fitting procedure indicates that the QP peak
does cross $E_F$, while at intermediate momenta it returns a gap smaller than at the antinodes
(Fig.\,\ref{fig2}f, right-hand side of cut II).

Let us now consider the momentum evolution of the QP lineshapes.
\begin{figure}[b!]
\centerline{\epsfig{figure=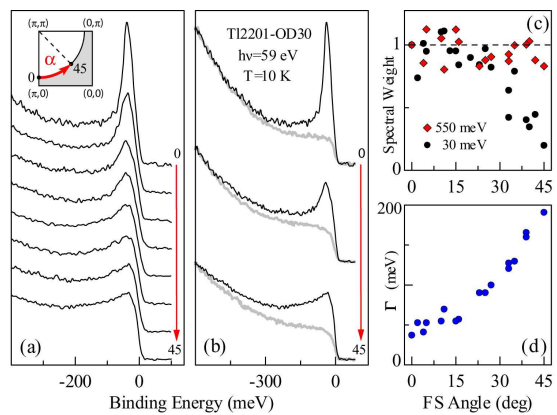,width=0.86\linewidth,clip=}} \vspace{-.1cm}
\caption{(color online). (a) Tl2201-OD30 ARPES spectra at $k$ slightly smaller than $k_F$ along the FS
contour (corresponding to a peak position of $\sim\!35$\,meV). (b) Selected spectra from (a) along with
corresponding $k\!\gg\!k_F$ background. (c,d) Spectral weight integrated over different energy ranges and QP
linewidth $\Gamma$ (see text for details) plotted vs. the FS-angle $\alpha$.} \label{fig3}
\end{figure}
A superficial inspection of the data in Fig.\,\ref{fig1}b,c shows that QPs are much broader in the nodal than
in the antinodal region. Furthermore, while in the nodal region the width of the QP peak increases as a
function of binding energy (Fig.\,\ref{fig1}b), as expected from phase space arguments, in the antinodal
region the sharpest peak is found at the bottom of the band (e.g., in Fig.\,\ref{fig2}e the linewidth
$\Gamma$ increases from $\sim\!30$ to 55\,meV when the QP peak disperses from $\sim\!39$ to 20\,meV). In
Fig.\,\ref{fig3}a we present a compilation of spectra taken along the FS contour but at $k$ slightly smaller
than $k_F$ and corresponding to a QP binding energy of $\sim\!35$\,meV (this choice being dictated by the
need to compare QP lineshapes not affected by the presence of a $d$-wave gap open along the FS and/or the
anomalous low-energy broadening shown in Fig.\,\ref{fig2}e). One can observe a sharp QP peak near ($\pi$,0),
which becomes progressively broader upon going towards ($\pi$/2,$\pi$/2). To qualitatively characterize the
evolution of the QP component of the spectral weight, we subtract from the ARPES spectra the background taken
from $k\!\gg\!k_F$ as in Fig.\,\ref{fig3}b \cite{Andrea:RMP,Kaminski}. Then we plot, as a function of the FS
angle, the integrated spectral weight normalized to the near-($\pi$,0) value (Fig.\,\ref{fig3}c) and the QP
linewidth $\Gamma$ estimated from the FWHM of the remaining QP component (Fig.\,\ref{fig3}d). The low-energy
spectral weight decreases monotonically along the FS in going from the antinodal to the nodal region and,
correspondingly, the linewidth increases from $\sim\!50$ to 180\,meV. However, the spectral weight integrated
over an energy window of $\sim\!550$\,meV is independent of the FS angle, so this seeming $k$-dependent loss
of QP coherence is intrinsic and is not simply a decrease of intensity due to, e.g., matrix element effects.

\begin{figure}[t!]
\centerline{\epsfig{figure=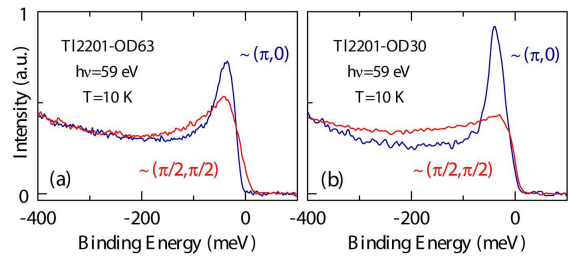,width=0.86\linewidth,clip=}} \vspace{-.1cm}
\caption{(color online). (a) Tl2201-OD63 and (b) Tl2201-OD30 spectra at $k\!\lesssim\!k_F$ in the nodal and
antinodal regions.}  \label{fig4}
\end{figure}

In order to put these observations into a broader context, we should recall that in underdoped cuprates QPs
are sharp near ($\pi$/2,$\pi$/2) and ill defined around ($\pi$,0), in the normal state. Upon increasing
doping, the antinodal QPs sharpen up, although they remain broader than the nodal QPs all the way to optimal
doping. Even in the SC state, in which case the QPs gain considerable coherence at all momenta and especially
in the antinodal region, the scattering rates determined by ARPES are still highly anisotropic with a minimum
at ($\pi$/2,$\pi$/2) \cite{Andrea:RMP}. At variance with this well established picture and the expectation
that the elementary excitations should become simply more isotropic upon overdoping, the SC state results
from Tl2201-OD30 show a reverse nodal/antinodal QP anisotropy (Fig.\,\ref{fig3} and\,\ref{fig4}b). A smaller
anisotropy of this sort is also observed on less overdoped Tl2201-OD63 (Fig.\,\ref{fig4}a), indicating a
trend with overdoping. In addition, recent ARPES data from overdoped $x\!=\!0.22$ LSCO \cite{Yoshida, Zhou}
are qualitatively comparable to those from Tl2201-OD63 (for both systems
$T_c\!\simeq\!\frac{2}{3}\,T_c^{max}$), suggesting that this behavior might be generic to overdoped cuprates.

Is this {\it QP anisotropy reversal} observed across optimal doping a signature of a quantum critical point
within the SC dome \cite{Tallon}? Interestingly, it was proposed that the proximity to a quantum phase
transition from a $d_{x^2-y^2}$ to a $d_{x^2-y^2}+id_{xy}$ superconductor as a function of some parameter,
possibly but not necessarily doping, may give rise to enhanced scattering of the gapless nodal QPs due to
their coupling to a low energy bosonic mode condensing at the phase transition \cite{Vojta}. However, while
an $id_{xy}$ pairing component leads to a state with no gapless fermionic excitations, the results from
Tl2201-OD30 do not show a gap in the nodal region. This, together with thermal conductivity experiments
\cite{Proust}, seems to indicate that the quantum criticality, if any, is not associated with the development
of an $id_{xy}$ pairing component or, perhaps, that doping is not the actual tuning parameter.

Alternatively, since a strongly $k$-dependent scattering rate is not supported by magnetotransport results
(as suggested by the small low-$T$ magnetoresistance, and by the comparison of low-$T$ resistivity and
cotangent of the Hall angle \cite{Mackenzie}), one may have to consider ARPES-specific QP broadening
mechanisms, such as elastic forward scattering or residual $k_z$ electronic dispersion. The latter may give
rise to $k_{||}$-dependent broadening of the ARPES features \cite{Sahrakorpi}, especially in overdoped Tl2201
for which a 0.4\% $c$-axis reduction with respect to optimal doping was reported \cite{Shimakawa} and a
three-dimensional coherent electronic behavior was observed \cite{Hussey}. However, the $c$-axis dispersion
vanishes at both nodal and antinodal points \cite{Hussey,Sahrakorpi}, which makes a connection between
observed QP anisotropy and finite $k_z$ dispersion not straightforward. Elastic small-angle (forward)
scattering, due to $n_i$ out-of-plane extended impurities such as cation substitution or interstitial oxygen,
contributes a term $\Gamma_{{\bf k}_F}\!\propto\!(n_iV_0^2)/(v_{{\bf k}_F}\kappa^3)$ to the total
normal-state electronic scattering (in the limit of large $\kappa^{-1}$, with $V_0$ and $\kappa^{-1}$ being
strength and range of the impurity potential \cite{Zhu}).
Above $T_c$, $\Gamma_{{\bf k}_F}$ is larger at the antinodes due to the smaller Fermi velocity $v_{{\bf
k}_F}$. Below $T_c$, however, due to energy conservation and the opening of the gap, $\Gamma_{{\bf k}_F}$ is
strongly suppressed in the antinodal but not in the nodal region, where small-angle scattering is still
pairbreaking \cite{Zhu}. This might provide an explanation for the anomalous $k$-dependence of the QP
scattering seen below $T_c$ on overdoped Tl2201. As for the doping dependence in Fig.\,\ref{fig4}, one might
speculate that while the broadening of the nodal QPs with doping is due to the increase of $n_i$ (i.e.,
oxygen content) and in turn of $\Gamma_{{\bf k}_F}$, the sharpening of the antinodal QPs reflects a decrease
of electronic correlations. On the other hand, the anomalous energy dependence of the ($\pi$,0) scattering
rate in Fig.\,\ref{fig2}e does not seem to be accounted for and might actually be indicative of a
distribution in gap magnitude due to electronic inhomogeneity, similar to what reported for Bi2212
\cite{Pan}.

In conclusion, overdoped Tl2201 is the first HTSC for which a surface-sensitive single-particle spectroscopy
and comparable bulk transport measurements have arrived at a quantitative agreement on a major feature such
as the normal state FS. As for the observed QP anisotropy reversal and its connection to various QP
broadening scenarios, including the recent proposal for an increase of antinodal lifetime through multiple
scattering off a single impurity \cite{Wakabayashi}, a more systematic study is required.

We thank K.M. Shen, D.G. Hawthorn, N.J.C. Ingle, N.E. Hussey, A.P. Mackenzie, M. Franz, D.J. Scalapino, and
G.A. Sawatzky for discussions. This work was supported by the CRC Program, NSERC, CIAR, and BCSI.


\vspace{-0.3cm}
\begin{references}
\vspace{-0.1cm}
\bibitem{Andrea:RMP}A. Damascelli {\it et al.}, Rev. Mod. Phys. {\bf 75}, 473 (2003).
\bibitem{Eisaki}H. Eisaki {\it et al.}, Phys. Rev. B {\bf 69}, 064512 (2004).
\bibitem{Proust}C. Proust {\it et al.}, Phys. Rev. Lett. {\bf 89}, 147003 (2002).
\bibitem{Tsuei}C.C. Tsuei {\it et al.}, Nature {\bf 387}, 481 (1997).
\bibitem{He}H. He {\it et al.}, Science {\bf 295}, 1045 (2002).
\bibitem{Hussey}N.E. Hussey {\it et al.}, Nature {\bf 425}, 814 (2003).
\bibitem{Peets}D.C. Peets, M.Sc. Thesis (UBC, 2003).
\bibitem{Yoshida}T. Yoshida {\it et al.}, Phys. Rev. Lett. {\bf 91}, 027001 (2003).
\bibitem{Zhou}X.J. Zhou {\it et al.}, Phys. Rev. Lett. {\bf 92}, 187001 (2004).
\bibitem{Shimakawa}Y. Shimakawa {\it et al.}, Phys. Rev. B {\bf 42}, 10\,165 (1990).
\bibitem{Hamann}D.R. Hamann {\it et al.}, Phys. Rev. B {\bf 38}, 5138 (1988).
\bibitem{Singh}D.J. Singh and W.E. Pickett, Physica C {\bf 203}, 193 (1992).
\bibitem{Mackenzie}A.P. Mackenzie {\it et al.}, Phys. Rev. B {\bf 53}, 5848 (1996).
\bibitem{NormanPRB}M.R. Norman, Phys. Rev. B {\bf 63}, 092509 (2001).
\bibitem{Jeff}J.D.F. Mottershead {\it et al.}, unpublished (2005).
\bibitem{NormanN}M.R. Norman {\it et al.}, Nature {\bf 392}, 157 (1998).
\bibitem{Kordyuk}A.A. Kordyuk {\it et al.}, Phys. Rev. B {\bf 67}, 064504  (2005).
\bibitem{Kaminski}A. Kaminski {\it et al.}, Phys. Rev. B {\bf 69}, 212509 (2004).
\bibitem{Tallon}J.L. Tallon and J.W. Loram, Physica C {\bf 349}, 53 (2001).
\bibitem{Vojta}M. Vojta {\it et al.}, Phys. Rev. Lett. {\bf 85}, 4940 (2000).
\bibitem{Sahrakorpi}S. Sahrakorpi {\it et al.}, cond-mat/0501500 (2005).
\bibitem{Zhu}L. Zhu {\it et al.}, Phys. Rev. B {\bf 70}, 214503  (2004).
\bibitem{Pan}S.H. Pan {\it et al.}, Nature {\bf 413}, 282 (2001).
\bibitem{Wakabayashi}K. Wakabayashi {\it et al.}, cond-mat/0504240 (2005).
\end{references}
\end{document}